\documentclass[letterpaper,twocolumn,superscriptaddress,floatfix]{revtex4-2}

\usepackage{inputenc}
\usepackage{bm}
\usepackage{multirow,amssymb,amsbsy,amsmath}
\usepackage{graphicx}
\usepackage{float}
\usepackage{verbatim}
\usepackage{color}
\makeatletter
\usepackage{pifont}
\usepackage{upgreek}
\usepackage{color}
\usepackage{indentfirst}
\usepackage{soul}
\usepackage{braket}
\usepackage[colorlinks=true,linkcolor=blue,citecolor=blue,urlcolor=blue]{hyperref}

\newcommand{\Rmnum}[1]{\expandafter\@slowromancap\romannumeral #1@}

\usepackage{caption}
\makeatother
\makeatletter 
\renewcommand{\section}{\@startsection{section}{1}{0mm}
	{-\baselineskip}{0.5\baselineskip}{\bf\leftline}}
\makeatother

\begin{document}
\title{Robust single divacancy defects near stacking faults in 4H-SiC under resonant excitation}
\affiliation{CAS Key Laboratory of Quantum Information, University of Science and Technology of China, Hefei, Anhui 230026, China}
\affiliation{CAS Center For Excellence in Quantum Information and Quantum Physics, University of Science and Technology of China, Hefei, Anhui 230026, China}
\affiliation{Hefei National Laboratory, University of Science and Technology of China, Hefei, Anhui 230088, China}
\affiliation{Institute of Advanced Semiconductors and Zhejiang Provincial Key Laboratory of Power Semiconductor Materials and Devices, ZJU-Hangzhou Global Scientific and Technological Innovation Center, Hangzhou, Zhejiang 311200, China}
\affiliation{State Key Laboratory of Silicon Materials and Advanced Semiconductors and School of Materials Science and Engineering, Zhejiang University, Hangzhou 310027, China}
\affiliation{College of Physics, Sichuan University, Chengdu, Sichuan 610065, China}
\affiliation{Center for Micro and Nanoscale Research and Fabrication, University of Science and Technology of China, Hefei, Anhui 230026, People's Republic of China}
\affiliation{State Key Laboratory of Functional Materials for Informatics, Shanghai Institute of Microsystem and Information Technology, Chinese Academy of Sciences, Shanghai 200050, China}

\author{Zhen-Xuan He}
\affiliation{CAS Key Laboratory of Quantum Information, University of Science and Technology of China, Hefei, Anhui 230026, China}
\affiliation{CAS Center For Excellence in Quantum Information and Quantum Physics, University of Science and Technology of China, Hefei, Anhui 230026, China}
\affiliation{Hefei National Laboratory, University of Science and Technology of China, Hefei, Anhui 230088, China}
   
\author{Ji-Yang Zhou}
\affiliation{CAS Key Laboratory of Quantum Information, University of Science and Technology of China, Hefei, Anhui 230026, China}
\affiliation{CAS Center For Excellence in Quantum Information and Quantum Physics, University of Science and Technology of China, Hefei, Anhui 230026, China}

\author{Wu-Xi Lin}
\affiliation{CAS Key Laboratory of Quantum Information, University of Science and Technology of China, Hefei, Anhui 230026, China}
\affiliation{CAS Center For Excellence in Quantum Information and Quantum Physics, University of Science and Technology of China, Hefei, Anhui 230026, China}
\affiliation{Hefei National Laboratory, University of Science and Technology of China, Hefei, Anhui 230088, China}	

\author{Qiang Li}
\affiliation{Institute of Advanced Semiconductors and Zhejiang Provincial Key Laboratory of Power Semiconductor Materials and Devices, ZJU-Hangzhou Global Scientific and Technological Innovation Center, Hangzhou, Zhejiang 311200, China}
\affiliation{State Key Laboratory of Silicon Materials and Advanced Semiconductors and School of Materials Science and Engineering, Zhejiang University, Hangzhou 310027, China}

\author{Rui-Jian Liang}
\affiliation{CAS Key Laboratory of Quantum Information, University of Science and Technology of China, Hefei, Anhui 230026, China}
\affiliation{CAS Center For Excellence in Quantum Information and Quantum Physics, University of Science and Technology of China, Hefei, Anhui 230026, China}

\author{Jun-Feng Wang}
\affiliation{College of Physics, Sichuan University, Chengdu, Sichuan 610065, China}
	
\author{Xiao-Lei Wen}
\affiliation{Center for Micro and Nanoscale Research and Fabrication, University of Science and Technology of China, Hefei, Anhui 230026, People's Republic of China}
 
\author{Zhi-He Hao}
\affiliation{CAS Key Laboratory of Quantum Information, University of Science and Technology of China, Hefei, Anhui 230026, China}
\affiliation{CAS Center For Excellence in Quantum Information and Quantum Physics,
University of Science and Technology of China, Hefei, Anhui 230026, China}

\author{Wei Liu}
\affiliation{CAS Key Laboratory of Quantum Information, University of Science and Technology of China, Hefei, Anhui 230026, China}
\affiliation{CAS Center For Excellence in Quantum Information and Quantum Physics, University of Science and Technology of China, Hefei, Anhui 230026, China}
  
\author{Shuo Ren}
\affiliation{CAS Key Laboratory of Quantum Information, University of Science and Technology of China, Hefei, Anhui 230026, China}
\affiliation{CAS Center For Excellence in Quantum Information and Quantum Physics, University of Science and Technology of China, Hefei, Anhui 230026, China}

\author{Hao Li}
\affiliation{State Key Laboratory of Functional Materials for Informatics, Shanghai Institute of Microsystem and Information Technology, Chinese Academy of Sciences, Shanghai 200050, China}

\author{Li-Xing You}
\affiliation{State Key Laboratory of Functional Materials for Informatics, Shanghai Institute of Microsystem and Information Technology, Chinese Academy of Sciences, Shanghai 200050, China}
   
\author{Jian-Shun Tang}
\affiliation{CAS Key Laboratory of Quantum Information, University of Science and Technology of China, Hefei, Anhui 230026, China}
\affiliation{CAS Center For Excellence in Quantum Information and Quantum Physics, University of Science and Technology of China, Hefei, Anhui 230026, China}
\affiliation{Hefei National Laboratory, University of Science and Technology of China, Hefei, Anhui 230088, China}
   
\author{Jin-Shi Xu}
\altaffiliation{Email: jsxu@ustc.edu.cn}
\affiliation{CAS Key Laboratory of Quantum Information, University of Science and Technology of China, Hefei, Anhui 230026, China}
\affiliation{CAS Center For Excellence in Quantum Information and Quantum Physics, University of Science and Technology of China, Hefei, Anhui 230026, China}
\affiliation{Hefei National Laboratory, University of Science and Technology of China, Hefei, Anhui 230088, China}

\author{Chuan-Feng Li}
\altaffiliation{Email: cfli@ustc.edu.cn}
\affiliation{CAS Key Laboratory of Quantum Information, University of Science and Technology of China, Hefei, Anhui 230026, China}
\affiliation{CAS Center For Excellence in Quantum Information and Quantum Physics, University of Science and Technology of China, Hefei, Anhui 230026, China}
\affiliation{Hefei National Laboratory, University of Science and Technology of China, Hefei, Anhui 230088, China}
 
\author{Guang-Can Guo}
\affiliation{CAS Key Laboratory of Quantum Information, University of Science and Technology of China, Hefei, Anhui 230026, China}
\affiliation{CAS Center For Excellence in Quantum Information and Quantum Physics, University of Science and Technology of China, Hefei, Anhui 230026, China}
\affiliation{Hefei National Laboratory, University of Science and Technology of China, Hefei, Anhui 230088, China}

\begin{abstract}
Color centers in silicon carbide (SiC) have demonstrated significant promise for quantum information processing. However, the undesirable ionization process that occurs during optical manipulation frequently causes fluctuations in the charge state and performance of these defects, thereby restricting the effectiveness of spin-photon interfaces. Recent predictions indicate that divacancy defects near stacking faults possess the capability to stabilize their neutral charge states, thereby providing robustness against photoionization effects. In this work, we present a comprehensive protocol for the scalable and targeted fabrication of single divacancy arrays in 4H-SiC using a high-resolution focused helium ion beam. Through photoluminescence emission (PLE) experiments, we demonstrate long-term emission stability with minimal linewidth shift ($\sim$ 50 MHz over 3 hours) for the single c-axis divacancies within stacking faults. By measuring the ionization rate for different polytypes of divacancies, we found that the divacancies within stacking faults are more robust against resonant excitation. Additionally, angle-resolved PLE spectra reveal their two resonant-transition lines with mutually orthogonal polarizations. Notably, the PLE linewidths are approximately 7 times narrower and the spin-coherent times are 6 times longer compared to divacancies generated via carbon-ion implantation. These findings highlight the immense potential of SiC divacancies for on-chip quantum photonics and the construction of efficient spin-to-photon interfaces, indicating a significant step forward in the development of quantum technologies.
\end{abstract}
	
  \maketitle
  
  \date{\today}
  
\section*{Introduction}

Silicon carbide (SiC) is a prominent example of third-generation semiconductors. It is renowned for its well-established inch-scale production, mature controlled-doping techniques, high-quality nanofabrication capabilities, and compatibility with CMOS-friendly processes. Similar to the nitrogen-vacancy (NV) centers in diamond, SiC also possesses a range of color centers, including silicon vacancies ($V_{Si}$) \cite{widmann2015coherent,nagy2018quantum,nagy2019high,Morioka2022SpinOpticalDA,fuchs2015engineering,simin2017locking}, divacancies (DVs) \cite{koehl2011room,falk2013polytype, christle2015isolated, christle2017isolated,Ivady2019EnhancedSO, Li2020RoomtemperatureCM}, NV centers \cite{Bardeleben2016NVCI, wang2020coherent, mu2020coherent}, and substitutional transition metal ions \cite{Diler2019CoherentCA, Wolfowicz2019VanadiumSQ, cilibrizzi2023ultra}. In comparison to NV centers in diamond that emit visible light, SiC color centers exhibit near-infrared photon luminescence (PL) with lower attenuation in fiber transitions. Moreover, the inherent response of SiC color centers to external fields makes them excellent platforms for magnetic, electric, thermal, and extreme-condition quantum sensing \cite{kraus2014magnetic,simin2015high,niethammer2016vector,yan2018coherent,zhou2017self,Wang2022MagneticDU,Liu2022CoherentCA,castelletto2023quantum,luo2023fabrication}. Leveraging well-established semiconductor integration technology, color centers can be manipulated within P-I-N junctions to achieve stark detuning \cite{miao2019electrically, Anderson2019ElectricalAO}, control charge states \cite{widmann2019electrical}, and generate electrically-driven single photon sources \cite{lohrmann2015single,Sato2018RoomTE}. Additionally, SiC color centers exhibit long spin-coherent times at both room and low temperatures \cite{ widmann2015coherent,simin2017locking,koehl2011room,christle2015isolated,christle2017isolated,Bourassa2020EntanglementAC}, as well as spin-selective optical transitions under resonant optical excitation \cite{christle2017isolated,miao2019electrically,Anderson2019ElectricalAO,nagy2018quantum,nagy2019high,Morioka2022SpinOpticalDA}, making them promising candidates for creating spin-to-photon interfaces. 

\begin{figure*}[htbp]
\centering
\includegraphics[scale = 0.6]{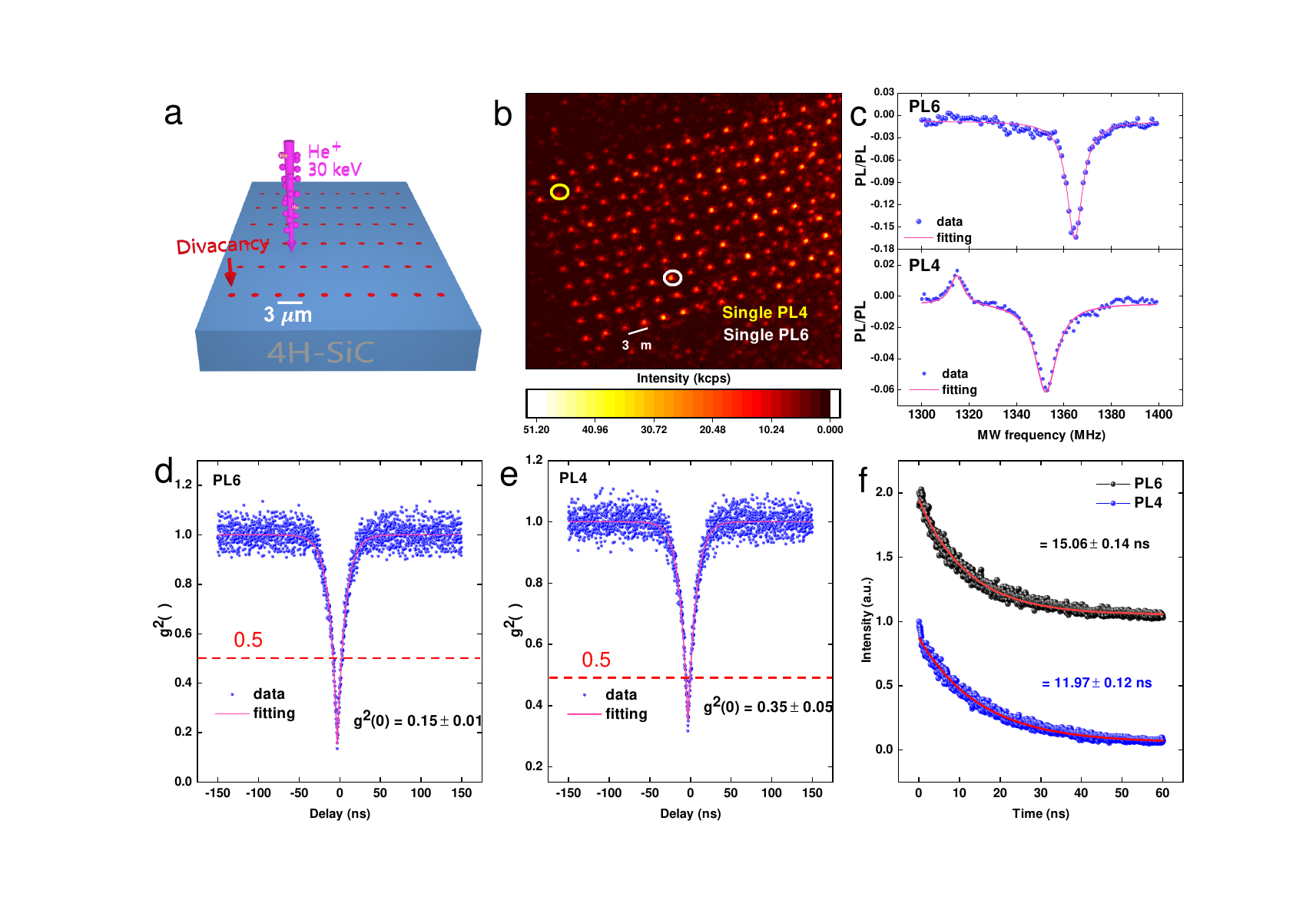}
  \caption{\textbf{FHIB-generated DV arrays in 4H-SiC and characterization of individual DVs.} (a) Schematic illustration depicting the fabrication of DV arrays in 4H-SiC utilizing FHIB. (b) Confocal scanning image displaying the DV arrays generated using a dose of 300 ions/spot. (c) Low-temperature ODMR spectrum obtained from the single white-circled PL6 (upper panel) and yellow-circled PL4 (lower panel) in (b). (d) As-measured second-order autocorrelation function for a single PL6. (e) As-measured second-order autocorrelation function for a single PL4.(f) The excited-state lifetimes for a single PL6 (black) and a single PL4 (blue).}
\label{Figure 1}
\end{figure*}

However, the charge-state instabilities of color centers under optical excitation have proven to be an important unresolved problem. Previous theoretical works have predicted that the quantum well structures in SiC, like stacking faults, which are of an extended defect, can stabilize the charge state of a point defect \cite{Ivady2019EnhancedSO}. Experimental works have also observed the stable photoemission under optical illumination of stacking-fault DVs in 4H-SiC (PL5 and PL6) ensembles \cite{wolfowicz2017optical}. Thus, stacking-faults DVs in 4H-SiC are potential high-optical-quality platforms for the quantum information process.

In this work, we investigated the application of a high-precise focused helium ion beam (FHIB) to create individual DV arrays in 4H-SiC. Through photoluminescence emission (PLE) experiments, we examined the optical properties under resonant excitation of the c-axis DV near stacking fault (PL6) and basal DV (PL4) in 4H-SiC. We found that the individual PL6 divacancies showed long-term emission stability for over 3 hours. Notably, by measuring the photoionization rates for PL4 and PL6, we found that stacking-fault-protected PL6 exhibits better resistance against resonant excitation-induced ionization. Additionally, we performed excitation angle-resolved PLE spectra measurements and observed two resonant-transition lines with mutually orthogonal polarizations, making them ideal for constructing spin-to-photon interfaces.  Moreover, our results showed a significant improvement in optical and spin properties for the helium ion implanted samples, with the PLE linewidths being 7 times narrower and the spin-coherent times being 6 times longer compared to divacancies generated via carbon-ion ($C^+$) implantation. Overall, our findings contribute to developing SiC-based quantum networking \cite{Xu2021silicon}, showing the potential of stacking-fault DV in this field.
  
  \section*{Results}
  \subsection{The FHIB-fabricated divacancy arrays}
  
  We utilized a helium ion microscope (HIM, ZEISS Orien NanoFab) to employ the FHIB technique for fabricating DV ensembles and arrays. The helium ion beam, with an incident energy of 30 keV, was accelerated and focused to achieve a spot size of approximately 0.5 nm. The implantation depth was simulated to be about 170 nm (see Supplementary Materials (SM) Fig. S1). The specific locations for each spot were determined based on a pre-designed layout. Fig. 1(a) illustrates the schematic diagram of the FHIB implantation process. DV ensembles were fabricated using the raster scanning mode with a dose of $1\times10^{15}$ ions/$cm^2$, while the arrays were created using the spot mode with varying doses (see  Materials and Methods and SM Section S1). Subsequently, the sample was annealed at 500 $ ^{o}$C for 2 hours, followed by a post-annealing at 900 $ ^{o}$C for 1 hour to facilitate DV formation.

  Fig. 1(b) displays the confocal scanning image of the generated DV arrays with an incident dose of 300 ions/spot, where a distance of 3 µm separated each spot. To investigate the polytypes of DVs within the array, we conducted measurements of the low-temperature (L-T) photoluminescence (PL) spectra and continuous-microwave (MW) optically-detected-magnetic-resonance (ODMR) spectra of both the ensembles and arrays at 4.5 K. (The L-T optical and spin-properties characterization are observable in Figs. S3 and S4 in SM) Within the array, we identified single PL6 defects. We have highlighted one representative PL6 and PL4 with a white and a yellow circle in Fig. 1(b), respectively. The L-T ODMR spectra for the circled PL6 (upper panel) and PL4 (lower panel) under a zero magnetic field ($B \approx 0$ G) are shown in Fig. 1(c). For the c-axis PL6, a single dip was observed at $\sim$ 1365 MHz, exhibiting an ODMR contrast of $\sim$ -0.16. This dip corresponds to the spin transition from the ground state $\ket{0}$ with the magnetic quantum number $m_{s}=0$ to the degenerate states $\ket{\pm1}$ with $m_{s}=\pm 1$ when $B \approx 0$ G \cite{koehl2011room}. By contrast, the basal PL4 exhibited one peak at $\sim$ 1316 MHz and one dip at $\sim$ 1353 MHz, representing the transitions from $\ket{0}$ to $\ket{-}$ ($=\frac{1}{\sqrt{2}}(\ket{1}-\ket{-1})$) and $\ket{+}$ ($=\frac{1}{\sqrt{2}}(\ket{1}+\ket{-1})$) in the ground state \cite{koehl2011room,miao2019electrically}. To confirm the single-photon emission properties of the defects, we conducted Hanbury Brown-Twiss (HBT) interference measurements. Figs. 1(d) and (e) illustrate the as-measured second-order autocorrelation functions $g^2(\tau)$ for PL6 and PL4 without background extraction. The fitted $g^2(0)$ values were 0.15 $\pm$ 0.01, and 0.35 $\pm$ 0.05, respectively, which are all far below 0.5, indicating single photon emission properties. Furthermore, we measured the PL spectrum to further confirm the polytype (see Fig. S4 in SM). The measured zero-phonon line (ZPL) positions were $\sim$ 1038 nm and $\sim$ 1078 nm, which were consistent with previous studies \cite{koehl2011room,Magnusson2018ExcitationPO,Li2020RoomtemperatureCM,miao2019electrically}. 
  We also measured the excited-state lifetime of the single PL6 and the single PL4, obtaining fitted values of $\tau=15.06 \pm 0.14$ ns and $\tau=11.97 \pm 0.12$ ns, as shown in Fig. 1(f).
  
  In addition, we also observed spots with ZPL emission and ODMR signals for PL4 with adjacent PL2. TABLE S3 and Fig. S4(d) in SM summarize the polytype of DVs for each spot within the 10 × 10 arrays shown in Fig. 1(b). The ODMR and PL signals were detected in 98 spots, with 82 spots exhibiting PL4, 11 spots showing PL4 with adjacent PL2 (denoted as PL2\&4), and 5 spots indicating PL6. Notably, the proportion of PL4 was the highest, consistent with the PL spectrum of the DV ensembles (see Fig. S2 in SM), where PL4 exhibited the strongest ZPL emissions compared to the other polytypes. The likelihood of finding DVs was significantly higher compared to electron-irradiated spots \cite{christle2017isolated}. While the PL5 spectrum was detected in the DV ensembles, no PL5 defects were encountered in the scanned 10$\times$10 array, suggesting a relatively limited efficiency in the preparation of PL5 defects.  

\subsection{Robust resonant excitation properties of PL6}

  The PLE experiments are important for investigating the spin-to-photon interface of solid-state color centers \cite{Babin_2021, Morioka2022SpinOpticalDA,nagy2019high,christle2017isolated}. We conducted PLE experiments on single PL6 and PL4 defects. We used a 914 nm off-resonant laser to initialize the spin state to $\ket{0}$ and perform deionization, followed by closing the off-resonant laser using a shutter. We then resonantly excited the DVs by scanning the wavelength of a tunable laser diode (ranging from 1030 nm to 1080 nm, encompassing the ZPLs of PL6 and PL4). We collected the PL intensity of phonon sideband (PSB) emission photons synchronously at each excitation wavelength, allowing us to obtain the responses to the varying excitation wavelengths.
  
\begin{figure*}[htbp]
\centering
\includegraphics[scale = 0.6]{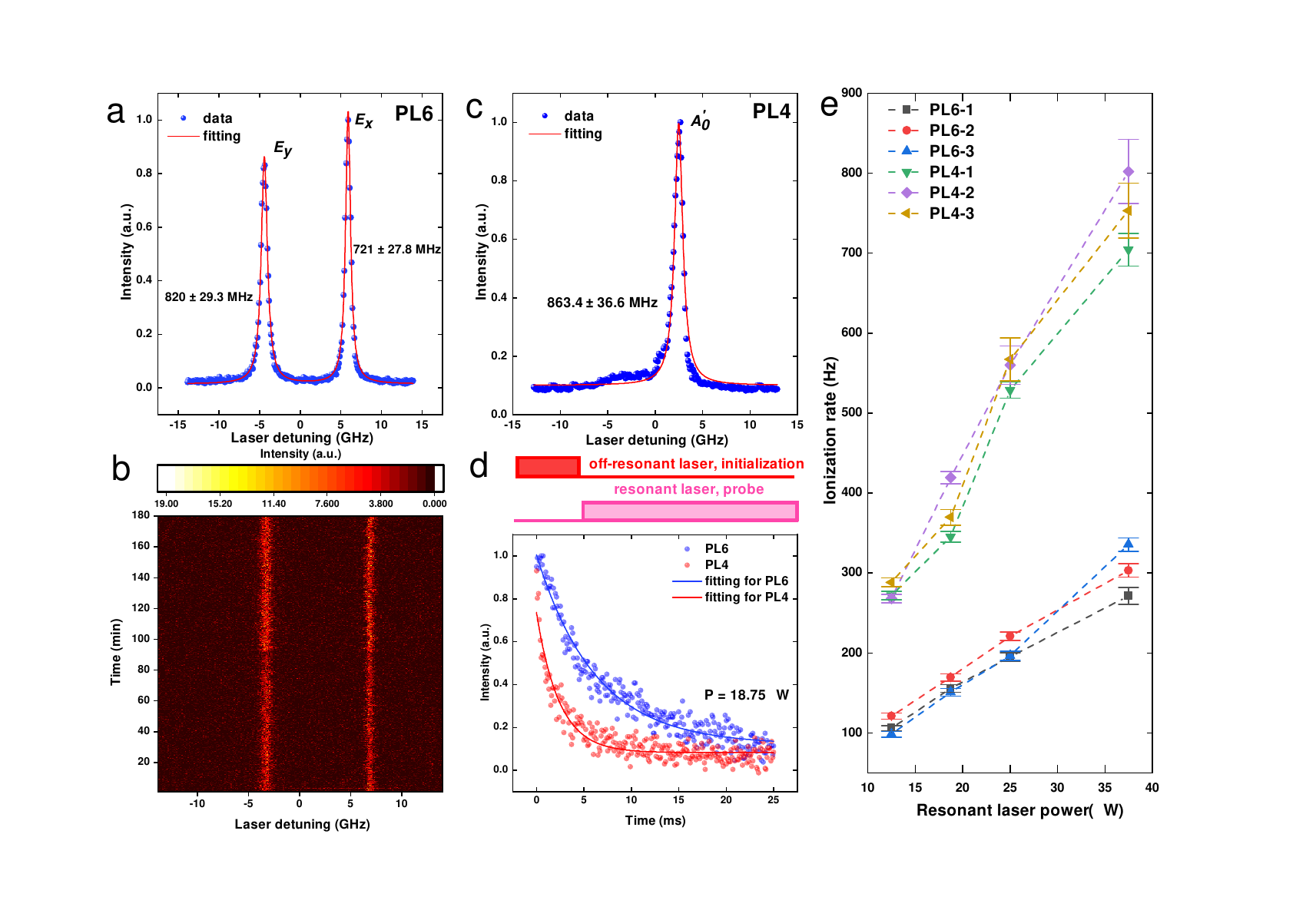}
  \caption{\textbf{Different resonant excitation properties for PL6 and PL4.} (a) PLE spectrum of a single PL6. The red line represents the Lorentz fitting. (b) Time-resolved PLE spectra of a single PL6 over 3 hours.  (c) PLE spectrum of a single PL4. The red line represents the Lorentz fitting. (d) Fluorescence decay of a single PL6 (blue dots) and a single PL4 (red dots) due to resonant excitation. The solid lines represent the exponential fittings. The upper panel presents the sequences to measure the data. (e) The ionization rates against resonant laser power for 3 single PL6 and 3 single PL4, showed good linear dependence. The ionization rate for PL4 is approximately 2.6 times higher than those of PL6.}
\label{Figure 2}
\end{figure*}

The PL6 defect has been previously identified as a c-axis defect \cite{koehl2011room, falk2013polytype, falk2014electrically}. In the case of a single PL6, we observed two resonant-transition lines originating from the ground state $\ket{0}$ to the excited states energy levels (ESELs). Fig. 2(a) displays the PLE spectrum of a single PL6, where the two branches correspond to the $E_x$ and $E_y$ transitions. The linewidths for the two branches were 721.22 $\pm$ 27.8 MHz and 820.03 $\pm$ 29.3 MHz, respectively. The zero laser detuning frequency is approximately 288.794 THz, which is about 1037.78 nm, in good agreement with the result of the PL spectrum. The linewidths observed in our measurements were approximately one order of magnitude larger than those obtained from the Fourier transformation of the measured lifetime. This phenomenon has been previously observed in similar studies of electron-irradiated DV defects \cite{christle2017isolated}, and it can be attributed to the interaction of lattice phonons induced by nitrogen impurities present in the commercial sample. Additionally, the insufficient implantation depth ($\sim$ 170 nm) could contribute to the broader line widths observed in our measurements. In comparison to electron-irradiated DV defects, which have an embedded depth of several micrometers, the FHIB-induced DV defects are slightly influenced by surface charge and electrical environments. By integrating the DV into a P-I-N junction, combined with charge depletion, it is possible to achieve a linewidth of approximately 20 MHz, approaching the lifetime limit \cite{Anderson2019ElectricalAO,miao2019electrically}. Furthermore, the linewidths were comparable to those of the substitutional Vandanium ions in 4H-SiC \cite{Wolfowicz2019VanadiumSQ} and electron-irradiated DVs in NIN-type SiC \cite{crook2020purcell}, being only around 4 to 8 times wider than the linewidths of electron-irradiated DVs (with a typical linewidth of 100 $\sim$ 200 MHz) in a commercial sample without charge depletion \cite{christle2017isolated,Anderson2019ElectricalAO}.

To test the stability of resonant excitation, we conducted a 3-hour analysis of the PLE spectra from a single PL6 defect. The time-resolved PLE spectra, shown in Fig. 2(b), exhibited remarkable stability in the positions of the two $E_x$ and $E_y$ transition lines during long-term monitoring. We performed double-peak Lorentz fittings for the acquired PLE spectra within the 3 hours and compared the fitted frequencies for both $E_x$ and $E_y$ with the central frequencies (see Fig. S5 in SM). The calculated average frequencies shift for $E_x$ and $E_y$ are $\pm$ 50.2 MHz and $\pm$ 51.3 MHz, comparable to the electron-irradiated DVs, whose long-term frequency shift is around 25 MHz \cite{Anderson2019ElectricalAO}. Such phenomena may originate from the protection of stacking-fault structures of PL6, which acts as a quantum well shelving. 

We proceeded with further characterization of the resonant excitation properties of the single PL4 defect. Fig  2(c) displays the PLE spectra for an individual PL4 defect, excited from the spin state $\ket{0}$ in the ground state, with a zero laser detuning frequency of approximately 278.105 THz. The ESELs associated with PL4 belong to the $C_{1h}$ point group, the excited states $A_0^{'}$ and $A_0^{"}$ are correlated with the ground $\ket{0}$ state \cite{miao2019electrically}. However, due to fast relaxation caused by the internal conversion from $A_0^{"}$ to $A_0^{'}$, only the $A_0^{'}$ state is observable in the experiment, which aligns with previous findings \cite{miao2019electrically}. The fitted linewidth was 863.4 $\pm$ 36.6 MHz, which was of the same magnitude as that of PL6.

The applied optical illumination drives undesired ionization processes that degrade the defects desired properties and functionality. For example, the fluorescence of color centers undergoes decay due to photoionization resulting from continuous resonant excitation. Therefore, it is crucial to identify color centers with robust emission properties that are resistant to ionization. To investigate the ionization process of the fabricated DVs, we employed a 10 µs off-resonant laser to initialize the DVs to the spin-0 ground state. Subsequently, a 30 ms continuous resonant laser pulse was applied, as depicted in the upper part of Fig. 2(d). The PL intensity during resonant excitation was synchronously collected and recorded using a counter.

\begin{figure*}[htbp]
\centering
\includegraphics[scale = 0.6]{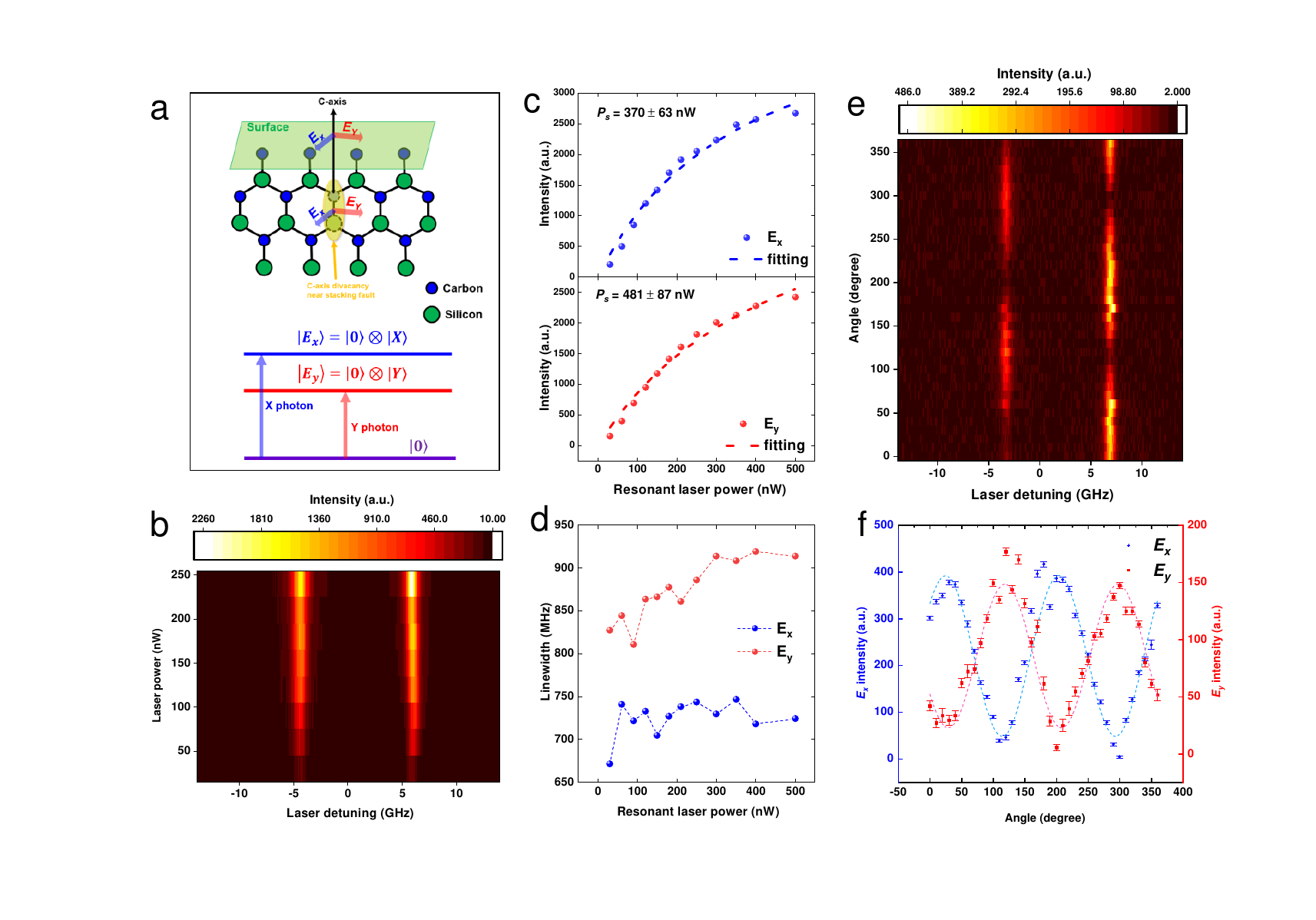}
\caption{\textbf{The optical polarization properties for $E_x$ and $E_y$ transitions} (a) Upper: the schematic diagram for the strain-split resonant excitation directions for the c-axis stacking-fault PL6. Lower: The schematic diagram illustrating the transitions of PL6 from the ground state (GS) $\ket{0}$ to the excited states (ES) $\ket{E_x}$ and $\ket{E_y}$. (b) Resonant-laser-power-resolved PLE spectra for a single PL6. (c) The PL intensity for the $E_x$ (blue dots, upper panel) and $E_y$ (red dots, lower panel) transitions against resonant laser power. The dash lines are the fitting. (d) The linewidths for $E_x$ (blue dots) and $E_y$ (red dots) of a single PL6 defect against resonant laser power. (e) Excitation angle-resolved PLE spectra. (f) Extracted data (blue dots for $E_x$ and red dots for $E_y$) from (e) with the corresponding cosine fitting.}
\label{Figure 3}
\end{figure*} 

Fig. 2(d) demonstrates the exponential decay of PL for both PL6 and PL4 under the same resonant excitation laser power ($P = 18.75$ µW). Notably, the PL intensity of a single PL4 decreases more rapidly than that of a single PL6, indicating a higher ionization rate $\Gamma$ for PL4 compared to PL6. The ionization rate $\Gamma$ was determined from the exponential fitting $I(t) = I_0 \exp(-\Gamma t)$, where $I(t)$ represents the time-resolved PL intensity under resonant excitation. To further verify this, we measured the ionization rates for three individual PL6 and three individual PL4 defects at different resonant laser powers ranging from approximately 12.5 µW to 37.5 µW, as shown in Fig. 2(e). Under a high resonant laser power of 37.5 µW, the ionization rates for both PL4 and PL6 were below 1 kHz. In contrast, for NV centers in diamond, the rate reaches 1 kHz at 10 µW \cite{robledo2010control}, confirming the greater resilience of DVs in 4H-SiC against photoionization. For each DV, the ionization rates exhibited a linear increase with the resonant driving power, with average linear parameters of 6.5 MHz/W for PL6 and 17.2 MHz/W for PL4, respectively. A previous experiment has reported a similar parameter of 10.2 MHz/W for c-axis DVs in 4H-SiC \cite{anderson2022five}. Notably, under the same resonant excitation laser power, the ionization rate for PL4 was approximately 2.6 times higher than that for PL6, confirming the superior resistance of PL6 defects to photoionization. These properties can be attributed to the quantum well effect arising from stacking faults. Previous works have predicted that a quantum well can lower the ionization energy of a point defect’s dark state so that the excitation laser will preferentially repopulate the point defect’s bright state \cite{Ivady2019EnhancedSO}. Another study has also demonstrated the robustness of stacking-fault DVs, as the PL intensity of PL6 remained stable under UV illumination \cite{wolfowicz2017optical}. This experiment provides direct evidence supporting the emission stability of single stacking-fault DVs under resonant excitation \cite{wolfowicz2017optical}.

\subsection{Optical polarization properties for PL6 under resonant excitation}

Alongside the robust optical properties under resonant excitation, we also observed distinguish polarization properties for PL6. From the group theory analysis, the ESELs of the c-axis DVs belong to a highly symmetrical $C_{3v}$ group due to the special lattice structure. The $E_x$ and $E_y$ transitions are split by a local strain ($E_{x(y)}$ is a direct product of $\ket{0}$ and X (Y) polarized photons, which are denoted as $\ket{X(Y)}$).  The excitation main axes for $E_x$ and $E_y$ exhibit mutual orthogonality, which are approximately parallel to the surface and perpendicular to the c-axis, as illustrated in Fig. 3(a). 

Firstly, we investigated the influence of resonant excitation laser power on both $E_x$ and $E_y$ transitions. Fig. 3(b) presents the power-resolved PLE spectra, demonstrating that the positions of $E_x$ and $E_y$ remained stable even as the laser power increased. The PL intensity against resonant laser power for $E_x$ and $E_y$ are observable in Fig. 3(c), both fitted by $I=I_s/(1+P/P_s)$ (dash lines), where $P_s$ represents the saturation resonant laser power. The fitted saturation powers are 370 $\pm$ 63 nW for $E_x$ and 481 $\pm$ 87 nW for $E_y$. Additionally, we performed linewidth fitting for PL6 under different excitation powers, as shown in Fig. 3(d). We didn't observe significant power-induced linewidth broadening. 

To confirm the polarization properties, we measured the excitation angle-resolved PLE spectra of the single PL6, which are presented in Fig. 3(e). By rotating the polarization angle of the resonant laser, the PL intensity of $E_x$ and $E_y$ varied, displaying a cosine relationship with the angle. The angles corresponding to maximum emission for $E_x$ and $E_y$ exhibited an offset of approximately 90 degrees, as depicted in Fig. 3(f) (data extracted from Fig. 3(e)). The results are in good agreement with group theory analysis, and a previous work observing strain-split two ZPL emission branches with mutually orthogonal polarization in c-axis DVs \cite{falk2014electrically}. These properties of PL6 resemble those of color centers with $C_{3v}$ symmetry, such as NV centers in diamond \cite{kaiser2009polarization,togan2010quantum}, PL1 and PL2 in 4H-SiC, and DVs in 3C-SiC \cite{Anderson2019ElectricalAO,christle2017isolated,falk2014electrically}, indicating their potential for constructing spin-photon entanglement through time-to-polarization conversion \cite{vasconcelos2020scalable}.

\begin{figure*}[htbp]
\centering
\includegraphics[scale = 0.6]{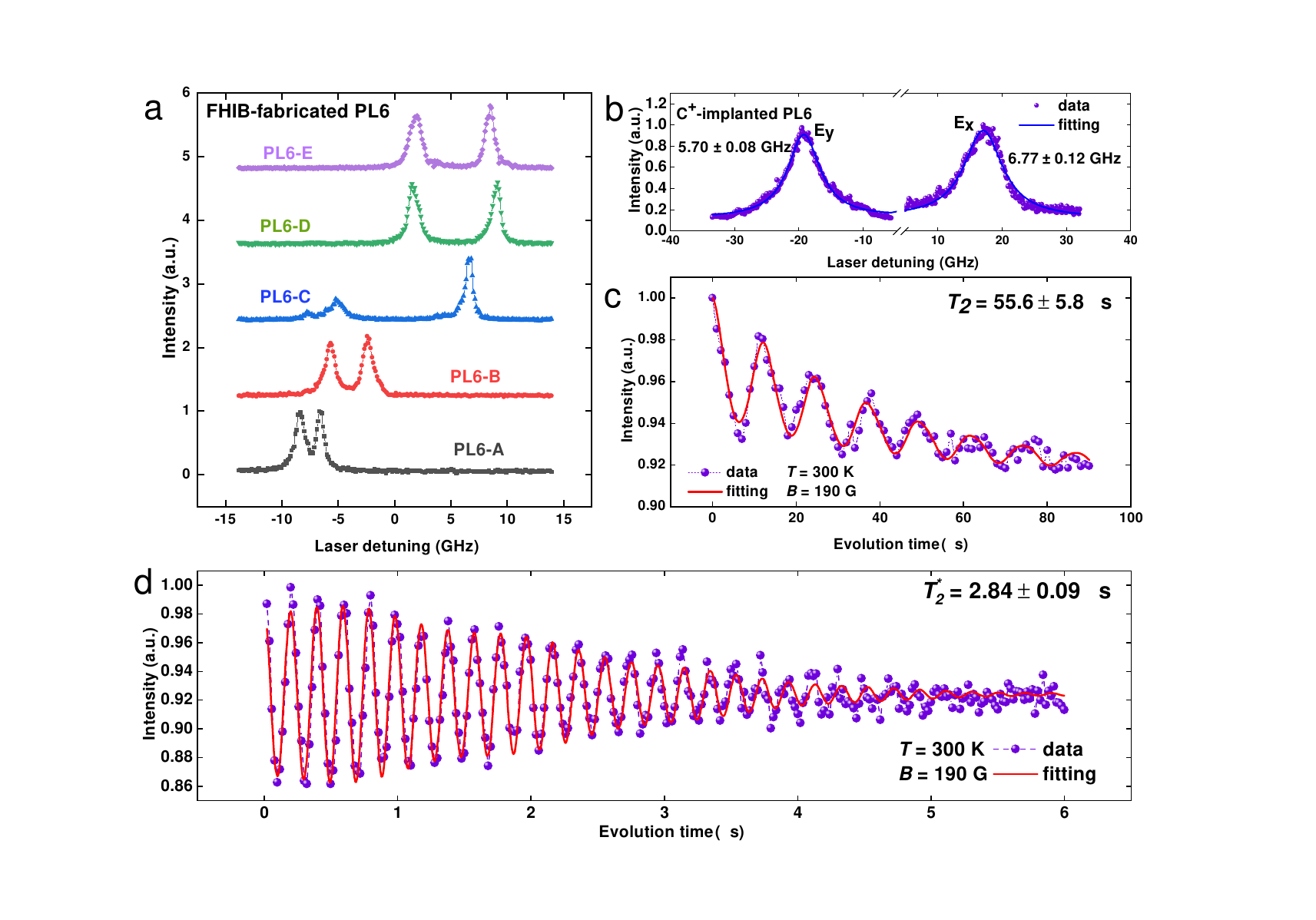}
\caption{\textbf{Comparision between FHIB-fabricated PL6 and $C^+$-implanted PL6.} (a) PLE spectra of 5 FHIB-fabricated individual PL6 defects. (b) PLE spectrum of a single PL6 generated by carbon-implantation. (c) The Hahn-echo measurement of for a single PL6 defect fabricated by FHIB. (d) Ramsey interference between the $\ket{0}$ and $\ket{-1}$ states for a single PL6 defect fabricated by FHIB.}
\label{Figure 4}
\end{figure*}

\subsection{Comparison between DVs fabricated by FHIB and $C^+$-implantation}

Furthermore, we conducted measurements of the PLE spectra (Fig. 4(a)) for another 5 single PL6 defects, labeled PL6-A to PL6-E. The linewidth values are summarized in Supplementary Material Table S4, and the average linewidths are approximately 879.3 MHz, consistent with the PL6 defect presented in Fig. 2. The time-resolved PLE spectra for these 5 defects, consisting of 80 cycles, can be found in Fig. S6, while the lifetime measurement results are shown in Fig. S7 in SM. For comparison, we also measured the PLE spectrum of a single PL6 defect generated through $C^+$ implantation (refer to Materials and Methods) \cite{Li2020RoomtemperatureCM}, as depicted in Fig. 4(b). The linewidths for the $E_x$ and $E_y$ transitions were 6.77 $\pm$ 0.12 GHz and 5.70 $\pm$ 0.08 GHz, respectively. In contrast, the linewidths of the DVs generated by FHIB were 7 to 8 times narrower compared to those generated by $C^+$ implantation. The observed linewidth broadening in carbon-implanted PL6 defects may be attributed to the shallower implantation depth in comparison to helium ion implantation (refer to Fig. S1 in SM), which exposes them to an unstable surface electrical environment during excitation, as well as the significant lattice damage caused by heavy ion implantation. Additionally, we compared the PLE linewidth between PL4 fabricated by FHIB and $C^+$-implantation (see Fig. S10 in SM), and the results were similar to those obtained for PL6.

We further conducted spin-coherent manipulations on the single PL6 defect illustrated in Fig. 2 at room temperature ($\sim$ 300 K). A c-axis magnetic field of approximately $190$ G was applied to separate the $\ket{\pm1}$ states in the ground states (refer to Fig. S8). We performed spin-coherent control between $\ket{0}$ and $\ket{-1}$. The Ramsey interference measurement was carried out with a microwave frequency detuning of $\Delta f = 6$ MHz, as shown in Fig. 4(d), resulting in a spin-coherent time of $T_2^* = 2.84 \pm 0.09$ $\mu$s. This value is comparable to electron-irradiated DVs in commercial 4H-SiC \cite{christle2015isolated} and approximately 6 times longer than that of $C^+$-implanted PL6 defects at room temperature, with $T_2^* \approx 450$ ns \cite{Li2020RoomtemperatureCM}. Fig. 4(c) illustrates the Hahn-echo measurements, exhibiting electron-spin-echo-envelope-modulation (ESEEM), with a fitted spin-coherent time of $T_2 = 55.6 \pm 5.8$ $\mu$s. This value is 2.5 times longer than that of $C^+$-implanted PL6 defects ($\sim$ 22 $\mu$s) \cite{Li2020RoomtemperatureCM}. We also examined the spin-coherent properties of two additional PL6 defects under the same experimental conditions, and their spin-coherent times were nearly identical (refer to Fig. S9 in SM). Furthermore, we performed spin-coherent control experiments on PL4 (refer to Fig. S11 in SM), and the coherent times were of a similar magnitude. 

The observed longer spin-coherent times can be attributed to the deeper helium ion implantation mentioned earlier, which protects against surface noise for the spin states. Similar properties have been reported for nitrogen-vacancy (NV) centers in diamonds, emphasizing the significance of deep implantation in preserving spin properties \cite{myers2014probing,sangtawesin2019origins}. Furthermore, the reduced lattice damage resulting from helium ion implantation may also contribute to the preservation of spin properties, as discussed in previous studies \cite{favaro2017tailoring}.

\section*{Discussion and conclusion}
In this study, we have investigated the feasibility of generating arrays of single divacancies using focused helium ion beam techniques. FHIB has shown great promise for precisely integrating color centers into nanophotonic structures, thanks to its high image resolution and generation efficiency. Moreover, the high-precision implantation provided by FHIB significantly enhances the coupling efficiency of defects compared to randomly distributed defects induced by electron irradiation. By studying the resonant excitation properties, we have observed that the stacking-fault DV PL6 exhibits long-term emission stability for over 3 hours. In comparison to the basal DV PL4, the photoionization rates for PL6 are 2.6 times lower, highlighting the significant role of stacking-fault protection. We have also discovered two mutually perpendicular resonant transition branches in PL6, similar to the NV centers in diamond. Furthermore, we have found that FHIB-induced DVs exhibit about 7 times narrower photoluminescence emission (PLE) linewidths and 6 times longer spin-coherent times compared to DVs generated by $C^+$ implantation. With its favorable spin-coherent properties, robust PLE emission with quantum-well protection, unique electron-spin-echo envelope modulation structures, and photon emission properties similar to NV centers in diamond, PL6 represents a promising candidate for constructing spin-photon entanglement networks based on time-to-polarization conversion and spin-photon entanglement states in ESELs \cite{togan2010quantum,vasconcelos2020scalable}. In conclusion, our findings highlight the potential of FHIB techniques and the unique properties of PL6 divacancies in silicon carbide for advancing the field of quantum information processing and constructing efficient spin-photon interfaces.

\section*{Materials and Methods}	
\textbf{Sample Fabrication}:
  We utilized a 4H-SiC sample with a 10 µm epitaxy layer. The helium ion microscope (HIM, Zeiss Orion NanoFab) with an imaging resolution of 0.5 nm and an incident energy of 30 keV was employed to implant the DV ensembles in scanning mode, with a dose of 1$\times$10$^{15}$ ions/cm$^2$. The DV arrays were fabricated using the spot mode with three different doses (400, 300, and 200 ions/spot). The sample was annealed at 500 $^{\circ}$C for 2 hours, followed by a post-annealing step at 900 $^{\circ}$C for 1 hour. The heating ramps for both annealing steps were set to 5 $^{\circ}$C/min.

  We further prepared a $C^+$-implanted sample. This sample was derived from the same wafer as the FHIB-fabricated sample. The array of single DVs and ensembles was generated by a 30-keV $C^+$ implantation using a pre-designed PMMA mask, followed by a 900 $^{\circ}$C annealing for 1 hour. The heating ramp during annealing was set to 5 $^{\circ}$C/min.

  \textbf{Room-temperature and Low-temperature Optical Setup}: For the room-temperature experiment, we employed a custom-built confocal laser scanning setup equipped with a piezo nanopositioner (E-727, PI). In the low-temperature experiment, a cryostat (Cryostation s200, Montana) fitted with an L-T piezo nanopositioner (ANC350, Attocube) was utilized to manipulate the sample position. To adjust the excitation laser trajectory and the photoluminescence (PL) of DVs, a fast steering mirror (Galvo, Thorlabs) was employed. The PL signal was collected through a single-mode fiber and detected by a superconducting nanowire single-photon detector (SNSPD, PHOTEC) with an approximate efficiency of 90\%.

  \textbf{Spin Control}: The microwave sequences were generated using a synthesized signal generator (MiniCircuits, SSG-6000 RC) and controlled by a switch (Mini-Circuits, ZASWA-2-50DR+). After amplification (Mini-Circuits, ZHL-25W-272+), the microwave signals were transmitted to a 50-$\mu$m-wide copper wire positioned above the surface of the 4H-SiC sample, within a distance of 80 $\mu$m from the DV arrays under investigation. The 914-nm continuous wave (CW) laser used for excitation was modulated using an acousto-optic modulator. The timing sequence of the electrical signals, which controlled and synchronized the laser, microwave, and counter, was generated by a pulse generator (SpinCore, PBESRPRO500).

  \textbf{PLE Experiment}: We employed an arbitrary function generator (AFG31252, Tektronix) to generate scanning voltages for the tunable laser diode (DL Pro, Toptica). The synchronization of the off-resonant laser, resonant laser, shutter, arbitrary function generator, and the counter was also controlled by a pulse generator (SpinCore, PBESRPRO500). 

  \textbf{Antibunching Experiment}: The coincidence correlation with variable delay times was measured using a time-to-digital converter (IDQ, ID800-TDC) with a resolution of 81 ps.

\section*{Supplementary Materials}
   This part contains sections S1 to S5, figures S1 to S11, and tables S1 to S5.

\section*{Author contributions} 
J.-S.X. and Z.-X.H. conceived the experiments. Z.-X.H. built the experiment set-up and performed the measurements with the help of J.-Y.Z., W.-X.L., Q.L., J.-F.W., Z.-H.H., W.L., R.S., R.-J.L., H.L., L.-X.Y. J.-S.T., and fabricated the sample with the help of Q.L. and X.L.W. Z.X.H.and J.-S.X. performed the data analysis and wrote the manuscript with contributions from all co-authors. J.-S.X., C.-F.L., and G.-C.G. supervised the project. All authors contributed to the discussion of the results.

\section*{Acknowledgement}	
  We thank Prof. Stefania Castelletto at the Royal Melbourne Institute of Technology for the helpful discussion. This work was supported by the Innovation Program for Quantum Science and Technology (Grants No. 2021ZD0301400), the National Natural Science Foundation of China (Grants No. U19A2075, No. 92365205, No. 11774335, No. 11821404 and No. 61725504), the Anhui Initiative in Quantum Information Technologies (Grant No. AHY060300). This work was partially performed at the University of Science and Technology of China Center for Micro and Nanoscale Research and Fabrication.
  
\section*{Competing interests}
	There are no conflicts to declare.
\section*{Data and materials availability}
	All data are available in the main text or the supplementary materials.

\end{document}